\documentclass[conference]{IEEEtran}
\IEEEoverridecommandlockouts
\usepackage[left=1.62cm,right=1.62cm,top=1.9cm]{geometry}
\usepackage{cite}
\usepackage[nolist]{acronym}
\usepackage[draft,bookmarks=false]{hyperref}
\usepackage{amsmath,amssymb,amsfonts}

\usepackage{algorithm}
\usepackage{algpseudocode}
\usepackage{graphicx}
\usepackage{textcomp}
\usepackage{orcidlink}
\usepackage{xcolor}

\usepackage[utf8]{inputenc}
\usepackage[T1]{fontenc}

\usepackage{mathtools}

\usepackage{booktabs}
\usepackage{siunitx}

\hypersetup{hidelinks}

\def\BibTeX{{\rm B\kern-.05em{\sc i\kern-.025em b}\kern-.08em
    T\kern-.1667em\lower.7ex\hbox{E}\kern-.125emX}}

\usepackage{listings}
\usepackage{color}
\usepackage{tikz}
\usetikzlibrary{arrows.meta,positioning,fit,shapes.multipart}

\tikzset{
  nf/.style={rectangle, rounded corners=2pt, draw, thick, align=center, minimum width=28mm, minimum height=8mm, fill=white},
  ext/.style={nf, dashed},
  data/.style={rectangle, draw, thick, align=center, minimum width=28mm, minimum height=8mm},
  call/.style={-{Stealth[length=2.5mm]}, thick},
  ref/.style={-{Stealth[length=2.0mm]}, dashed, gray},
  lbl/.style={font=\footnotesize, inner sep=1pt, outer sep=1pt}
}

\lstdefinelanguage{json}{
    basicstyle=\ttfamily\small,
    numbers=none,
    showstringspaces=false,
    breaklines=true,
    frame=lines,
    backgroundcolor=\color[gray]{0.95},
    morestring=[b]",
    literate=
     *{0}{{{\color{black}0}}}{1}
      {1}{{{\color{black}1}}}{1}
      {2}{{{\color{black}2}}}{1}
      {3}{{{\color{black}3}}}{1}
      {4}{{{\color{black}4}}}{1}
      {5}{{{\color{black}5}}}{1}
      {6}{{{\color{black}6}}}{1}
      {7}{{{\color{black}7}}}{1}
      {8}{{{\color{black}8}}}{1}
      {9}{{{\color{black}9}}}{1}
      {:}{{{\color{black}:}}}{1},
}

\newcommand{\vect}[1]{\boldsymbol{#1}}

\acrodef{3gpp}[3GPP]{3rd Generation Partnership Project}
\acrodef{5GS}{5G System}
\acrodef{6gs}[6GS]{6G System}
\acrodef{af}[AF]{Application Function}
\acrodef{agv}[AGV]{Automated Guided Vehicles}
\acrodef{amf}[AMF]{Access and Mobility Management Function}
\acrodef{aoa}[AOA]{Angle of Arrival}

\acrodef{bs}[BS]{Base Station}

\acrodef{cn}[CN]{Core Network}

\acrodef{dt}[DT]{Digital Twin}

\acrodef{etsi}[ETSI]{European Telecommunications Standards Institute}

\acrodef{gdpr}[GDPR]{General Data Protection Regulation}
\acrodef{gnb}[gNB]{Next-Generation NodeB}

\acrodef{IETF}{Internet Engineering Task Force}
\acrodef{isac}[ISAC]{Integrated Sensing and Communication}
\acrodef{isg}[ISG]{Industry Specification Group}

\acrodef{json}[JSON]{JavaSrcipt Object Notation}

\acrodef{kpi}[KPI]{Key Performance Indicator}

\acrodef{lmf}[LMF]{Location Management Function}

\acrodef{ml}[ML]{Machine Learning}

\acrodef{nf}[NF]{Network Function}
\acrodef{nwdaf}[NWDAF]{Network Data Analytics Function}
\acrodef{pcf}[PCF]{Policy Control Function}
\acrodef{ran}[RAN]{Radio Access Network}
\acrodef{rcs}[RCS]{Radar Cross Section}
\acrodef{rf}[RF]{Radio Frequency}

\acrodef{spf}[SPF]{Sensing Processing Function}
\acrodef{sf}[SF]{Sensing Function}
\acrodef{stid}[STID]{Sensing Task ID}
\acrodef{sba}[SBA]{Service-Based Architecture}
\acrodef{scf}[SCF]{Sensing Coordination Function}
\acrodef{srx}[SRx]{Sensing Receiver}
\acrodef{stx}[STx]{Sensing Transmitter}

\acrodef{trp}[TRP]{Transmission Reception Point}

\acrodef{ue}[UE]{User Equipment}
\acrodef{sdp}[SDP]{Sensing Data Point}
\acrodef{se}[SE]{Sensing Entity}
\acrodefplural{se}{Sensing Entities}
\acrodef{ssc}[SSC]{Sensing Service Consumer}
\acrodef{sdsf}[SDSF]{Sensing Data Storage Function}
\acrodef{uav}[UAV]{unmanned aerial vehicle}
\acrodef{sdo}[SDO]{standards development organization}
\begin{document}
\title{Architectural Enhancements for Efficient Sensing Data Utilization in 6G ISAC
\thanks{This work has been partially funded by the European Commission Horizon Europe SNS JU projects MultiX (GA 101192521).}}

\author{\IEEEauthorblockN{Muhammad Awais Jadoon, Sebastian Robitzsch}
\IEEEauthorblockA{InterDigital Europe Ltd, London, United Kingdom,\\
Email: \{Muhammad.AwaisJadoon, Sebastian.Robitzsch\}@interdigital.com}
}

\maketitle

\begin{abstract}
Current architecture proposals within \aclp{sdo} such as ETSI and 3GPP enable sensing capabilities in mobile networks; however, they do not include a repository for storing sensing data. Such a repository can be used for AI model training and to complement ongoing sensing service provisioning by improving efficiency and accuracy. One way of realizing this is through the fusion of historical sensing data with live sensing data. In this paper, we study historical and live sensing data fusion for \acl{isac} in future 6G systems and introduce a \acl{sdsf} to store historical sensing data and sensing results. We show how the \acl{sdsf} can be used with other network functions in a 6G architecture proposition for \acl{isac}. We validate our proposal with a measurement model and show performance improvements in terms of detection probability and false-alarm rate. The network functionality to fuse and process sensing data combines \emph{live} sensing measurements with previously sensed historical sensing data using a map-aware hard filter that rejects detections consistent with known static structures. Our simulation illustrates that, for a traffic junction scenario, map-aware hard filtering substantially reduces false alarms without degrading detection probability.
\end{abstract}


\section{Introduction}
\ac{isac} aims to endow mobile networks with environmental perception in addition to legacy communications. Recent and ongoing 3GPP efforts \cite{3gpp22870} and \ac{etsi} \ac{isg} \ac{isac} \cite{ETSIISC001} work define requirements and use cases for \ac{isac} as well as architectural enhancements required to enable \ac{isac} in 6G systems. Across all proposed solutions for 5G-A in 3GPP \cite{3GPP22837} and 6G in ETSI \cite{ETSIISC003}, a new \ac{sf} is proposed in the 6G \ac{cn} to coordinate sensing tasks and process sensing data. In many ETSI proposed solutions, the \ac{sf} is further divided into a \ac{scf} and a \ac{spf}. \ac{scf} coordinates sensing tasks or sensing operations while the \ac{spf} processes sensing measurements/data and generates sensing results. Sensing results are then provided to the \ac{ssc}. However, there is no standardized function responsible for persistent storage of sensing data and results. We motivate the need for a \ac{sdsf} to retain historical sensing data that can be reused to reduce redundant live re-sensing, bandwidth, and energy consumption, while preserving requested sensing service \acp{kpi}. Historical sensing data refers to the sensing data that is collected previously using sensing operations from both 3GPP and non-3GPP sources. 

\begin{figure}[htb]
    \centering
    \includegraphics[width=\linewidth]{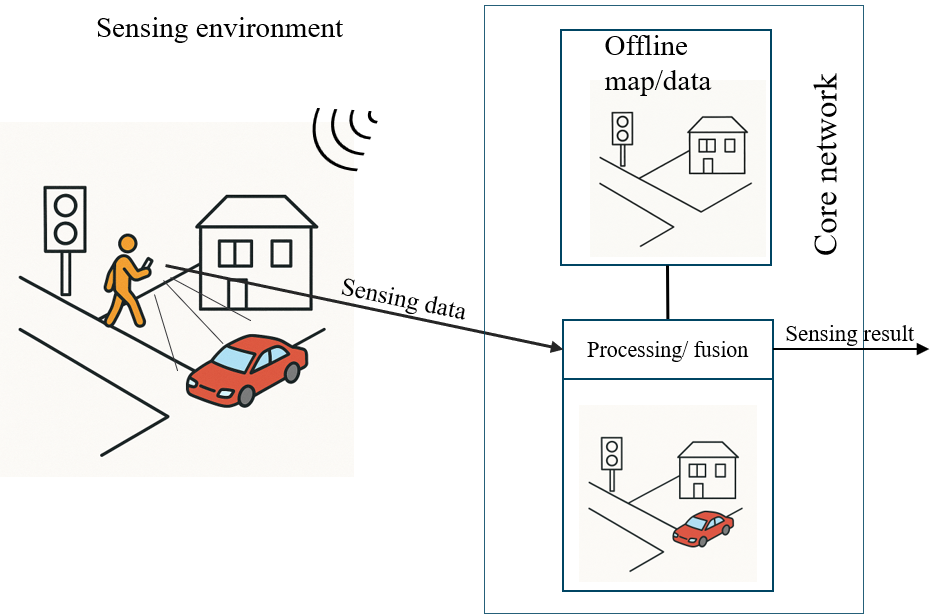}
    \caption{System Model}
    \label{fig:sys_model}
\end{figure}

An example use case to illustrate where such functionality can significantly improve the offered sensing service is provided in \figurename~\ref{fig:sys_model}. The figure depicts a vehicular use case where a traffic junction is sensed and the 6G system is requested to provide sensing results to a third-party \ac{ssc} informing about any moving environmental object and its object type (e.g. pedestrian, vehicle, lorry, cyclist, motorcyclist).

In this paper, we propose an \ac{sdsf} as a new 6G \ac{cn} NF to store sensing data/results and report or expose them to \acp{nf} or third-party \acp{af}, respectively. We use a map-aware hard filter at \ac{spf} that uses historical sensing data maps to reject static-consistent returns. We provide a measurement model to validate our proposal. We show that the historical sensing data (map) can reduce clutter, and a successful fusion reduces false-alarm rate, while probability is not affected significantly as compared to when only live sensing data is used.

The rest of the paper is structured as follows. In Section~\ref{sec:relatedwork}, we provide related works on a potential 6G-\ac{isac} architecture. In Section~\ref{sec:proposd_arch}, we present the proposed architectural enhancements to support storage of sensing data. In Section~\ref{sec:measmodel}, we present how historical sensing data can be used with live sensing. We present performance metrics in Section~\ref{sec:metrics} that we use to validate our results. Simulation results are discussed in Section~\ref{sec:results}. Finally, we provide conclusions in Section~\ref{sec:conclusion}. 

\section{Related Work}\label{sec:relatedwork}
ETSI's ISG ISAC published 6G use cases on \ac{isac} \cite{ETSIISC001} ETSI ISG ISAC GR001 \cite{ETSIISC001} also reports detailed deployment considerations. Meanwhile, 3GPP started studying 6G use cases \cite{3GPP22837} themselves, structured around all six usage scenarios in ITU-R's IMT 2030 vision framework with one being \ac{isac}.

3GPP is currently moving to Stage 2 work in 5G-A to enable mono-static gNB-only sensing \cite{3GPP22137, 3GPP22837}. These developments led to the discussion of a sensing-enabled 5G architecture, which already provides the flexibility to support new functionalities and services such as \ac{isac} and the enablement of \ac{uav} use cases in 5G. The diverse 6G \ac{isac} use cases studied in \ac{etsi} and \ac{3gpp}, together with their functional and performance requirements, necessitates the evolution of existing 5G system capabilities and, most likely, the introduction of new 6G \ac{cn} \acp{nf}. Accordingly, several research efforts and standardization activities are underway to determine how \ac{5GS} should evolve and how the 6G architecture may incorporate additional functionalities to support emerging \ac{isac} use cases.\cite{gersing2024architectureproposal6gsystems, hexa-x-ii-d2.3, Moro_icc24, robitzsch2025architectureconsiderationsisac6g, MultiX-D2.1}.

In \cite{Liu23_arch, 11230762}, architecture for \ac{isac} is proposed, where a dedicated \ac{scf} is used to coordinate sensing activities. The functionality of an \ac{scf} contains configuration parameters for the sensing request. The sensing request contains information regarding the type of the target and also the \ac{kpi} requirements. A new architecture has been proposed recently in \cite{gersing2024architectureproposal6gsystems} where a dedicated \ac{scf} serves as an anchor to all sensing task-related functions and where its services can be used by other \acp{nf}. Similarly, in \cite{Lyazidi_24, jadoon_icc26}, a comparable architecture is proposed, comprising two entities: one responsible for coordination, i.e., \ac{scf} and configuration related to the sensing task, and the other dedicated to processing the sensing data, i.e., \ac{spf}. In \cite{jadoon_vtc26}, we had proposed an efficient way of collecting sensing data. However, only \cite{MultiX-D2.1} calls out a new 6G \ac{cn} \ac{nf} providing storage capabilities, but without any detailed procedures on how to leverage an \ac{sdsf} and the reuse of sensing data to efficiently utilize sensing data to avoid redundant sensing and saving energy and resources.

\section{Proposed Architectural Enhancements in 6G \ac{isac}}\label{sec:proposd_arch}
\subsection{Proposed Functions in 6GS}
\begin{figure}
    \centering    
    \includegraphics[width=\linewidth] {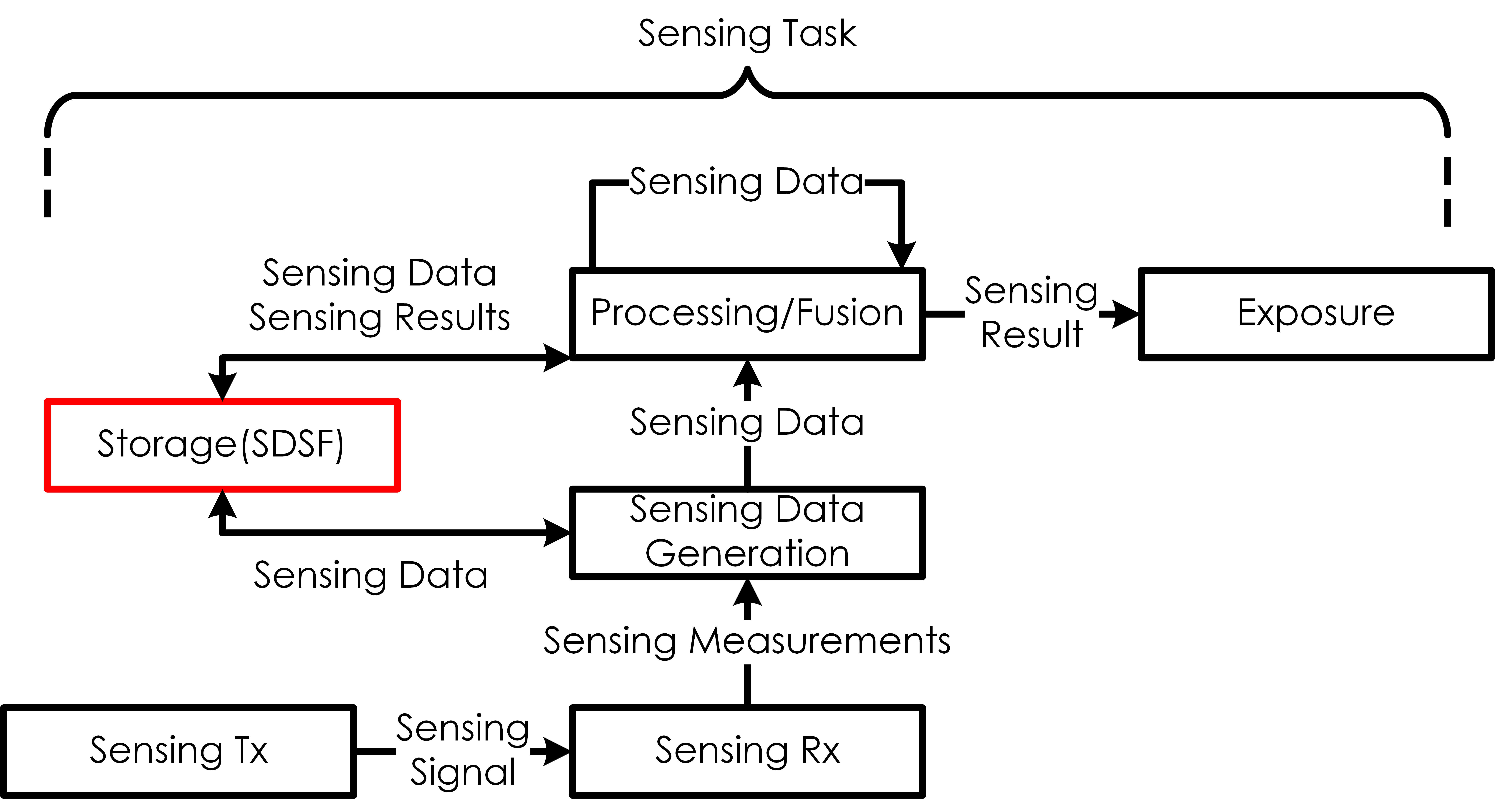}
    \caption{Proposed Sensing Workflow for 6GS with \ac{sdsf} as storage function}
    \label{fig:sensing_workflow}
\end{figure}

A new logical \ac{nf}, called \ac{sdsf}, is proposed as shown in Fig~\ref{fig:sensing_workflow}, that stores sensing data and/or sensing results in a standardized format (e.g., point clouds, spatio-temporal maps) indexed by a \ac{stid}, time and context. \ac{sdsf} supports query/subscribe interfaces for other \acp{nf} in the architecture such as the \ac{spf}.

The \ac{scf} orchestrates sensing tasks and configures \acp{se} (e.g., \ac{ue} and/or \ac{gnb}) based on the input received from the \ac{spf}. The \ac{spf} maps sensing service \ac{kpi} requirements to sensing measurements that are required to achieve the aforementioned \acp{kpi}. These sensing measurement determinations are provided to the \ac{scf} to configure \acp{se}. 

The \ac{spf} collects sensing data, either from gNB as sensing data generation entity or from \acp{se}, and is capable of retrieving historical sensing data from the \ac{sdsf}. In such scenario, the \ac{spf} performs the fusion of live and historical sensing data and outputs sensing results. The fusion task can be performed using advanced signal processing algorithms or AIML models. How these algorithms operate is outside of the scope of this paper. For validity of the proposed storage function, we provide a measurement model that uses map-aware filtering, detailed in Section~\ref{sec:measmodel}.

For simplicity, we assume that \ac{scf} and \ac{spf} are part of the \ac{sf}.

In the context of the vehicular use case, the \ac{sdsf} enables the reuse of static context (i.e., offline sensing data) and reduces measurement load/bandwidth in the \ac{spf} without re-sensing the entire scene. The \ac{scf} can decide when offline data suffices and when to refresh the map if sensing service \acp{kpi} drift.

\subsection{Proposed Solution for Historical Sensing Data Utilization for a Sensing Task in 6GS}
Stepwise procedure:
\begin{enumerate}
    \item \acp{se} register their sensing capabilities with the \ac{sf}.
    
    \item The \acf{ssc} sends a sensing service request to the \ac{sf}.  
    The request includes \ac{kpi} requirements for the sensing service, the consent status for using historical sensing data, and optionally the allowed age/freshness of historical data.  
    An \ac{ssc} may be a \ac{ran} node requiring sensing analytics (e.g., beamforming recommendations, energy optimization guidance), an application on a \ac{ue}, another \ac{nf}, or a third-party \ac{af}.
    
    \item The \ac{sf} determines whether the \ac{kpi} requirements can be satisfied, sends an acknowledgment to the \ac{ssc}, and appends the \ac{stid} associated with the request.
    
    \item In parallel, the \ac{sf} sends a sensing data usage policy request to the \ac{pcf}.  
    The request may include: consent tokens for \acp{se}, the purpose of sensing data usage (e.g., RAN analytics, sensing results generation), the location/area of the sensing task, and any charging policy information.

\begin{figure}
    \centering   
    \includegraphics[width=\linewidth] {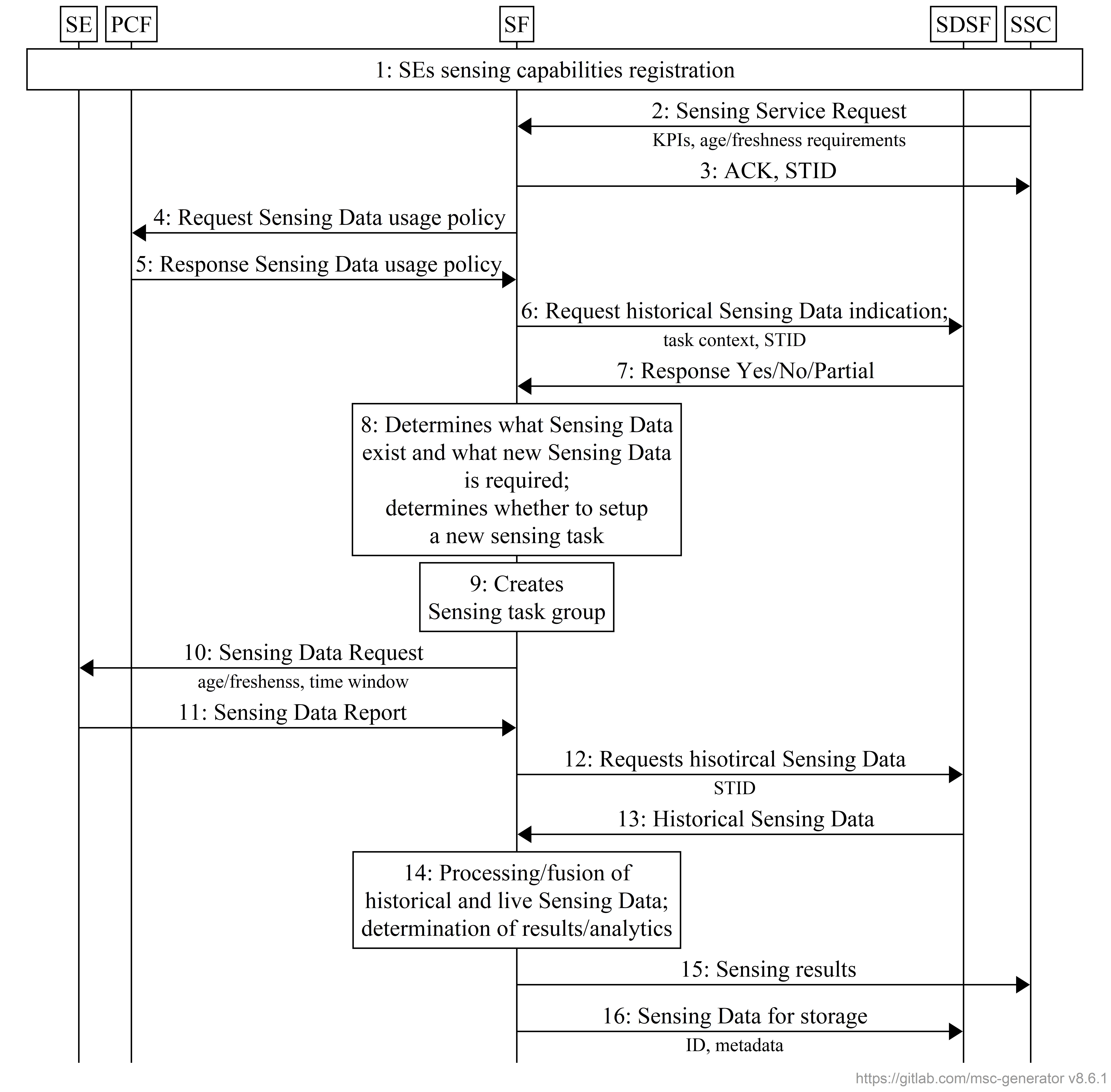}
    \caption{Call flow for the proposed \ac{sdsf}}
    \label{fig:sensing_quality}
\end{figure}

    \item The \ac{pcf} processes the request and returns policies associated with the \ac{stid}.  
    The \ac{pcf} may permit, deny, or permit with obligations.  
    It may provide consent revocation rules, charging rules, and lists of prohibited areas.  
    These policies assist the \ac{sf} in determining whether historical sensing data can be used when generating sensing results or analytics for the RAN.
    
    \item After receiving the policies from the \ac{pcf}, the \ac{sf} queries the \ac{sdsf} for an indication of whether the required sensing data (raw, processed, or high-level) exists in the \ac{sdsf}'s storage.  
    The \ac{sf} uses the sensing task context (e.g., target type, weather conditions, location, time) and the \ac{stid}, enabling the \ac{sf} and \ac{sdsf} to exchange historical sensing data specific to the task.
    
    \item The \ac{sdsf} responds with an indication of whether the required historical sensing data exists, does not exist, or partially exists.  \label{step:x}
    In the partial case, the \ac{sdsf} provides details about which portions are available and which are missing.
    
    \item Based on the indication from the \ac{sdsf}, the \ac{sf} determines what new or live sensing data is required to fulfill the sensing service \ac{kpi} requirements.  
    If live sensing data is needed, the \ac{sf} initiates the setup of a new sensing task by selecting \acp{se} to perform sensing and report sensing data.
    
    \item If live sensing data is required, the \ac{sf} creates a sensing task group composed of one or more \acp{se} responsible for generating and reporting sensing data.
    
    \item The \ac{sf} sends sensing data requests to the \acp{se} in the sensing task group.  
    The request includes sensing configuration parameters~\cite{jadoon_icc26}, such as which sensing data elements are required, the time window of interest, reporting behavior (periodic or one-shot), and age/freshness constraints.  
    This enables \acp{se} to determine whether existing buffered sensing data is still fresh enough to be reused.
    
    \item The \acp{se} perform sensing and send live sensing data to the \ac{sf}. 
    If the \acp{se} possess buffered sensing data that satisfies the freshness constraints, this data is also reported.
    
    \item Once sufficient live sensing data is collected, the \ac{sf} requests the historical sensing data from the \ac{sdsf} that was indicated as available in Step~\ref{step:x}.
    
    \item The \ac{sdsf} sends the corresponding historical sensing data to the \ac{sf}.
    
    \item The \ac{sf} fuses or combines the live sensing data from the \acp{se} with the historical sensing data from the \ac{sdsf} to generate sensing results that meet the KPI requirements of the \ac{ssc}.  
    If the \ac{ssc} is a \ac{ran} node, the \ac{sf} may also derive sensing analytics to support \ac{ran} resource and energy optimization.
    
    \item If the fused and processed sensing data satisfy the sensing service \ac{kpi} requirements, the \ac{sf} sends the sensing results to the \ac{ssc}.
    
    \item The \ac{sf} sends updated or newly generated sensing data to the \ac{sdsf} for storage.  
    The \ac{sf} uses identifiers such as the \ac{stid} along with metadata (e.g., storage location, data aging policy, context changes) to ensure the data is correctly archived and managed.
\end{enumerate}

In the following section, we develop a  measurement model that enables us to examine how historical sensing data can be fused with live sensing data and thereby validate the proposed enhancements to 6G \ac{isac} architecture. 

\section{Measurement Model}
\label{sec:measmodel}
In this Section, we formalize how a local polar measurement $(r,\beta)$, produced by a \ac{se}, is expressed in the world (Cartesian) frame and how its uncertainty is propagated for use in filtering and association. 
Consider a 2‑D planar junction (horizontal plane). An \ac{se} $s$, e.g., a UE or BS with location and orientation of $(x_s,\,y_s,\,\theta_s)$ reports a local polar measurement obtained from sensing signals as
\begin{align}
\label{eq:z_def}
\mathbf{z}
=
\begin{bmatrix} r \\ \beta \end{bmatrix}
+ \mathbf{n},
\qquad
\mathbf{n}\sim \mathcal{N}\!\big(\mathbf{0},\,\Sigma_{rb}\big),
\quad
\Sigma_{rb}=\mathrm{diag}(\sigma_r^2,\,\sigma_\beta^2),
\end{align}

where $r\in\mathbb{R}_{>0}$ is \emph{range} (meters), i.e., distance from the \ac{se} to the detected target; $\beta\in(-\pi,\pi]$ is \emph{bearing} (radians), i.e., the azimuth angle in the SE’s local frame relative to the \ac{se} boresight (forward axis). {In world (global) azimuth, the direction is $\beta+\theta_s$ \cite{Skolnik2001,Thrun2005}. $\mathbf{n}$ is the zero‑mean measurement noise, independent across $r$ and $\beta$ with variances $\sigma_r^2$ and $\sigma_\beta^2$ (m$^2$, rad$^2$).

These measurements are obtained in local coordinates of the \ac{se} in order to fuse measurements from several \acp{se} with different orientations, \ac{spf} converts the detection to global frame as,

\begin{align}
\label{eq:polar_to_cart_full}
\begin{bmatrix} x \\[2pt] y \end{bmatrix}
=
\begin{bmatrix} x_s \\[2pt] y_s \end{bmatrix}
+
R(\theta_s)
\begin{bmatrix} r\cos\beta \\[2pt] r\sin\beta \end{bmatrix}, \\
\end{align}
where
\begin{align}
R(\theta_s)=
\begin{bmatrix}
\cos\theta_s & -\sin\theta_s \\
\sin\theta_s & \ \cos\theta_s
\end{bmatrix}.
\end{align}
The vector $\big[r\cos\beta,\ r\sin\beta\big]^T$ is the detection’s Cartesian coordinates in the \ac{se}'s local frame.
This is the standard radar/robotics transform used for multi‑sensor perception and map alignment \cite{Skolnik2001,Thrun2005}.

The \ac{spf} needs $\Sigma_{xy}$ to perform validation gating, i.e., decide if a detection is consistent with a static map. The transform in \eqref{eq:polar_to_cart_full} is nonlinear in $(r,\beta)$ and to express the positional uncertainty of a target detection, we use a first‑order Extended Kalman-filter linearization around the current measurement mean \cite{Thrun2005} as:
\begin{align}
\label{eq:jacobian_J}
J
=\frac{\partial(x,y)}{\partial(r,\beta)}
=
\begin{bmatrix}
\cos\beta & -r\sin\beta \\
\sin\beta & \ \ r\cos\beta
\end{bmatrix},
\end{align}
and therefore,
\begin{align}
\Sigma_{xy}
=
J\,\Sigma_{rb}\,J^\top,
\end{align}
where $J$ is the Jacobian of the $(r,\beta)\!\to\!(x,y)$ mapping and it quantifies how small changes in $r$ or $\beta$ perturb $(x,y)$. $\Sigma_{xy}$ is Cartesian covariance (m$^2$). This means that at long range $r$, the lateral standard deviation is approximately $r\,\sigma_\beta$, while the radial deviation is $\sigma_r$.

\subsection{Detection Gating}
Gating is performed at the \ac{spf} and it is the process of deciding whether a detection matches a target. Once each detection has its Cartesian covariance, we can now check whether it is statistically consistent with a predicted target or with a known static map. With $\hat{\mathbf{x}}=(x,y)$ from \eqref{eq:polar_to_cart_full}, \ac{spf} accepts a detection for target $i$ at $\mathbf{x}_i$ if it falls within a \emph{validation gate}:
\begin{equation}
\label{eq:det_gate}
\min_j \big\|\hat{\mathbf{x}}_j - \mathbf{x}_i\big\| \le g_{\text{det}},
\end{equation}
where $g_{\text{det}}$ is an Euclidean gate (in meters). 
We use Euclidean distance for simplicity. An alternative is Mahalanobis‑distance gating with $\Sigma_{xy}$, which is used in classical tracking \cite{BarShalom1988,BarShalom1995}.

\subsection{Map‑Aware Filtering}
For \emph{map‑aware hard filtering}, we reject a detection if it lies inside the \emph{dilated} static map region:

\begin{equation}
\label{eq:hard_mask}
\hat{\mathbf{x}}\in \mathcal{M}\oplus \mathcal{B}(g),
\end{equation}

where $\mathcal{M}$ is the union of static structures (static map with global coordinates), $\oplus$ is the Minkowski sum, and $\mathcal{B}(g)$ is a disk of radius $g$ (meters). The hard mask rejects any detections whose back-projected $(x, y)$ lies within the dilated map. This means that every true measurement, if it happens close to a building edges, may be rejected. 

\section{Performance Metrics}
\label{sec:metrics}
We focus on two sensing metrics: \emph{detection probability} ($P_d$) and \emph{false-alarm rate} ($\overline{\mathrm{FA}}$). Both of these metrics directly quantify how well a system separates true targets from clutter at a given operating point (gate), and they are the standard event-level quantities used in validation gating and association \cite{BarShalom1988,BarShalom1995}.

$P_d$ (true-positive probability) measures target observability and association success under a given gate, while $\overline{\mathrm{FA}}$ (false positives per step) measures the residual clutter that survives filtering and gating.

$P_d$ and $\overline{\mathrm{FA}}$ are standard metrics to validate that historical sensing data (e.g., maps) can be fused with live sensing to suppress clutter while maintaining detection performance. They align with classical gating/association theory \cite{BarShalom1988,BarShalom1995} and admit a clean analytical expectation under Poisson thinning.

At time $t$, let $\{\hat{\vect{x}}_j(t)\}$ be the set of accepted detections in the global (world) frame after any pre-processing (e.g., map-aware hard mask), and let $\vect{x}_i(t)$ be the true position of target $i$.
We use Euclidean validation gate with radius $g_{\text{det}}$ (meters) for metric computation (consistent across baselines).
Symbols: $i\in\{1,\ldots,N\}$ indexes targets; $j$ indexes detections; $t\in\{1,\ldots,T\}$ indexes time steps; $\|\cdot\|$ is the Euclidean norm.

\subsection{Detection Probability}
\label{subsec:metrics_pd}
A detection for target $i$ occurs at time $t$ if at least one accepted detection falls within the gate:
\begin{equation}
\label{eq:det_event_metrics}
\mathcal{D}_i(t)\ \triangleq\ \Big(\min_{j}\ \big\|\hat{\vect{x}}_j(t)-\vect{x}_i(t)\big\|\ \le\ g_{\text{det}}\Big),
\end{equation}

where $\mathcal{D}_i(t)$ is the detection event (true/false), $\hat{\vect{x}}_j(t)$ is the $j$‑th accepted detection in world coordinates, $\vect{x}_i(t)$ is the true target position and $g_{\text{det}}$ is the gate radius (m).

\begin{algorithm}[!t]
\caption{Map-aware Filtering}
\label{alg:mapawarefilterig}
\begin{algorithmic}[1]
\For{each time step $t$}
    \State Collect raw detections from sensing entities (live data).
    \If{historical + live fusion is enabled}
        \State Apply map-aware hard mask using $\mathcal{M}$ to remove detections in $\mathcal{M}\oplus\mathcal{B}(g)$.
    \EndIf
    \State Back-project accepted detections to world coordinates.
    \State For each in-bounds target $i$, perform the detection test (Eq.~\eqref{eq:det_event_metrics}) and update $P_{d,i}$ (Eq.~\eqref{eq:pd_def}).
    \State Count unmatched detections as false-alarms (Eq.~\eqref{eq:fa_step}) and accumulate $\overline{\mathrm{FA}}$ (Eq.~\eqref{eq:fa_avg}).
\EndFor
\State Compute final metrics $P_d$ and $\overline{\mathrm{FA}}$ over all time steps.
\State \Return $P_d,\,\overline{\mathrm{FA}}$
\end{algorithmic}
\end{algorithm}
Let $\mathcal{T}_i=\{t:\ \vect{x}_i(t)\ \text{is inside the sensing area}\}$ with $T_i=|\mathcal{T}_i|$.
The per‑target and average detection probabilities respectively are
\begin{equation}\label{eq:pd_def}
P_{d,i}=\frac{1}{T_i}\sum_{t\in\mathcal{T}_i}\mathbf{1}\!\left[\mathcal{D}_i(t)\right],
\end{equation}
\begin{align}\label{eq:pd_avg}
    P_d=\frac{1}{N}\sum_{i=1}^{N} P_{d,i}.
\end{align}
where $\mathbf{1}[\cdot]$ is the indicator function, $N$ is the number of targets and $P_d$ is the averaged detection probability across targets.

$P_d$ increases if (i) the target has higher per‑step detection success or (ii) association is easier (less clutter within the gate). Map-aware filtering primarily aids (ii) by removing static-consistent returns near structures, thereby reducing mismatches.

\subsection{False-alarm rate}
\label{subsec:metrics_fa}
A detection is a \emph{false-alarm} at time $t$ if it does not match any target within the gate:
\begin{equation}
\label{eq:fa_step}
\mathrm{FA}_t=\#\Big\{j:\ \min_{i} \big\|\hat{\vect{x}}_j(t)-\vect{x}_i(t)\big\| > g_{\text{det}}\Big\}.
\end{equation}

$\mathrm{FA}_t$ is the number of unmatched detections at step $t$ and $\#\{\cdot\}$ counts elements.

The \emph{average} false-alarms per step is
\begin{equation}
\label{eq:fa_avg}
\overline{\mathrm{FA}}=\frac{1}{T}\sum_{t=1}^{T} \mathrm{FA}_t,
\end{equation}
where $T$ is the total number of steps. $\overline{\mathrm{FA}}$ measures residual clutter that survived filtering and the gate.

\section{Simulation Results and Discussion}\label{sec:results}
\subsection{Setup}
We simulate an experiment with a \SI{120}{\meter}\,$\times$\,\SI{120}{\meter} junction with two rectangular buildings acting as a static map structure $\mathcal{M}$. Two \acp{se} are placed at $(0,0)$ and $(120,0)$ with $\theta_s=0$ to collect range-bearing measurements with additive Gaussian noise ($\sigma_r=\SI{0.8}{\meter}$, $\sigma_\beta=\ang{2}$). Each \ac{se} detects a target with fixed probability $P_{\text{det}} = 0.95$. We generate $N=8$ moving targets (mixed horizontal/vertical paths) per frame, while clutter is generated as a Poisson process with mean $\lambda_{FA}=60$ detections/step (across \acp{se}), with $70\%$ concentrated near building edges (Gaussian jitter), $30\%$ uniform background. 

For each mask margin $g \in [0, 5] \text{m}$ and validation gate $g_\text{det} \in \{1, 2, 3, 4, 5, 10\}$, detections are first filtered by map-aware hard mask and then associated with true targets using the Euclidean gate $g_\text{det}$ as shown in Algorithm~\ref{alg:mapawarefilterig}. 
\begin{figure}
    \centering
    \includegraphics[width=\linewidth]{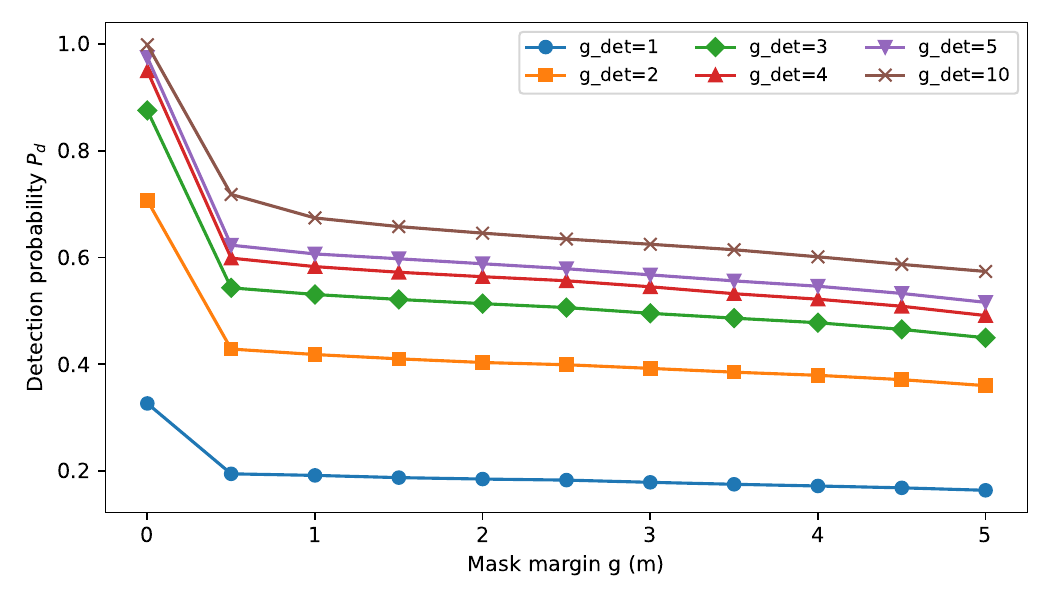}
    \caption{Probability of Detection $P_d$ $g$ for different values of $g$ and $g_\text{det}$}
    \label{fig:Pd_vs_g}
\end{figure}

\begin{figure}
    \centering
    \includegraphics[width=\linewidth]{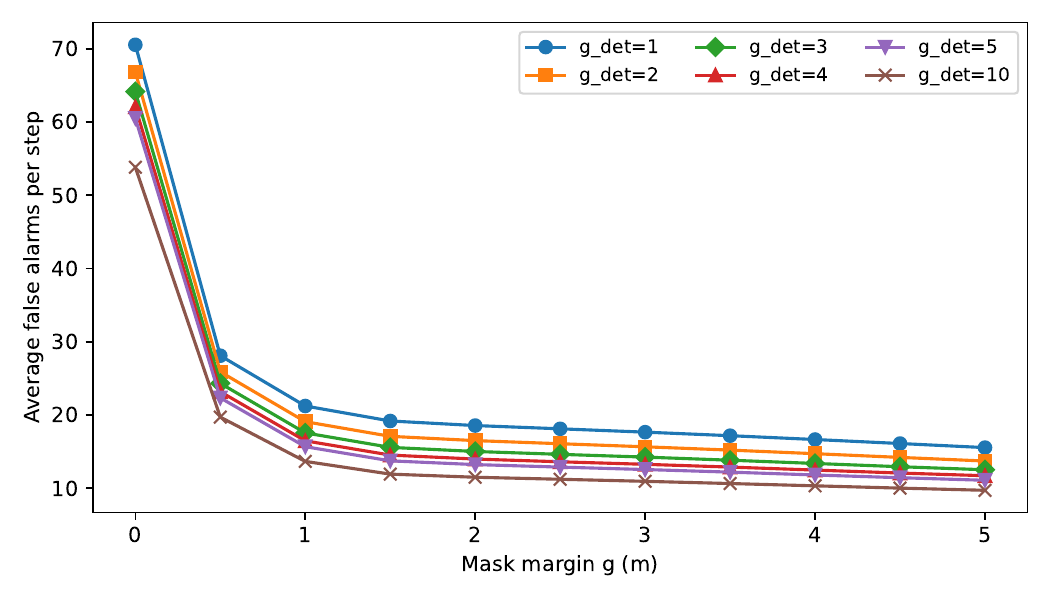}
    \caption{Average False-Alarm Rate for different values of $g$ and $g_\text{det}$}
    \label{fig:FA_vs_g}
\end{figure}

\subsection{Results Discussion}
We present simulation results for $P_d$ and $\overline{\mathrm{FA}}$, averaged over 50 Monte Carlo realizations and shown in \figurename~\ref{fig:Pd_vs_g} and \figurename~\ref{fig:FA_vs_g}, respectively.
In \figurename~\ref{fig:Pd_vs_g}, when $g=0$, it means all detections are considered and hence the $P_d$ is highest for this case. As $g$ is increased, we observe a dip in $P_d$. This is because the hard mask rejects any detections whose back-projected $(x, y)$ lies within the dilated map. This means that every true measurement, if it happens close to a building edges, may be rejected. Therefore, if a real moving target happens to pass within that buffer $g (.)$ around static structures, its noisy measurements may land inside the mask and be discarded, even though the object itself is valid. That directly reduces $P_d$. In reality, the correct use of historical sensing data would be probabilistic or conditional: e.g.,
``Reduce confidence for static regions but still allow moving objects through if motion evidence exists.''
The hard-mask is just a simplified implementation that prioritizes false-alarm reduction over completeness. The monotonic increase of $P_d$ with $g_\text{det}$ agrees with gating theory, i.e., a relaxed gate reclassifies near-misses as successful detections.

\figurename~\ref{fig:FA_vs_g} demonstrates a strong and monotonic reduction in false-alarms as $g$ grows. When $g=0$, all clutter, particularly reflections concentrated around static edges, is passed to the association stage, yielding the highest FA count. Expanding the mask margin gradually removes detections geometrically consistent with the static map, realizing a Poisson-thinning effect on clutter. For 
$g \geq 2, \overline{\mathrm{FA}}$ drops by roughly 70–80 \% relative to the live-only baseline. Increasing 
$g_\text{det}$ further lowers the measured $\overline{\mathrm{FA}}$ because unmatched detections become less likely when the gate is wide; however, in a full tracker this may correspond to occasional false associations. Therefore, there is tradeoff in selecting $g$ and $g_\text{det}$ and must be chosen carefully. 

\section{Conclusion}\label{sec:conclusion}
We motivated historical and live sensing data fusion for \ac{isac} and introduced the \ac{sdsf} in a 6G architecture proposition to store sensing data/results. We validated our proposal using a map-aware hard filter at the \ac{spf} (part of the \ac{sf}). A controlled evaluation shows large false-alarm reductions with no loss in detection probability. The formulation is grounded in standard range--azimuth models, validation gating, and Poisson thinning. Future work includes uncertainty-aware margins. We used a fixed margin $g$ for hard masking. A covariance-aware margin $g(r,\beta)=k\sqrt{\lambda_{\max}(\Sigma_{xy})}$ can be used along with 3-D geometry and more realistic maps of the environment. One future direction can be to analyze how much energy and resources are saved using historical sensing data.

\bibliographystyle{IEEEtran}
\bibliography{references} 

@book{Skolnik2001,
  author    = {Merrill I. Skolnik},
  title     = {Introduction to Radar Systems},
  edition   = {3rd},
  publisher = {McGraw-Hill},
  year      = {2001}
}

@book{Thrun2005,
  author    = {Sebastian Thrun and Wolfram Burgard and Dieter Fox},
  title     = {Probabilistic Robotics},
  publisher = {MIT Press},
  year      = {2005}
}

@book{BarShalom1988,
  author    = {Yaakov Bar-Shalom and Thomas E. Fortmann},
  title     = {Tracking and Data Association},
  publisher = {Academic Press},
  year      = {1988}
}

@book{BarShalom1995,
  author    = {Yaakov Bar-Shalom and Xiao-Rong Li},
  title     = {Multitarget-Multisensor Tracking: Principles and Techniques},
  publisher = {YBS Publishing},
  year      = {1995}
}

@misc{3GPP22137,
  author       = {{3GPP}},
  title        = {{TS} 22.137: Service Requirements for Integrated Sensing and Communication (Stage 1), {Rel}-19},
  year         = {2024},
  howpublished = {\url{https://www.3gpp.org}}
}

@misc{3GPP22837,
  author       = {{3GPP}},
  title        = {{TR} 22.837: Feasibility Study on Integrated Sensing and Communication, Rel-19},
  year         = {2024},
  howpublished = {\url{https://www.3gpp.org}}
}

@misc{3gpp22870,
    organization = {3GPP},
    title = {{TR} 22.870: Study on 6{G} Use Cases and Service Requirements, Release 20},
    url = {https://www.3gpp.org/ftp/Specs/archive/22_series/22.870/22870-101.zip}
}

@misc{ETSIISC001,
  author       = {{ETSI ISG ISAC}},
  title        = {{GR ISC} 001 V1.1.1 (2025-03): Integrated Sensing and Communications ({ISAC}); Use Cases and Deployment Scenarios},
  year         = {2025},
  url ={https://www.etsi.org/deliver/etsi_gr/ISC/001_099/001/01.01.01_60/gr_ISC001v010101p.pdf}
}

@misc{MultiX-D2.1,
organization = {MultiX Consoritum},
title = {D2.1: MultiX perceptive 6G-RAN system design v1 – focus on initial DASH design},
year = {2025},
url = {https://multix-6g.eu/wp-content/uploads/2025/12/D2.1.pdf}
}

@misc{robitzsch2025architectureconsiderationsisac6g,
      title={Architecture Considerations for ISAC in 6G}, 
      author={Sebastian Robitzsch and Laksh Bhatia and Konstantinos G. Filis and Neda Petreska and Michael Bahr and Pablo Picazo Martinez and Xi Li},
      year={2025},
      eprint={2508.13736},
      archivePrefix={arXiv},
      primaryClass={cs.NI},
      url={https://arxiv.org/abs/2508.13736}, 
}

@misc{ETSIISC003,
  orgianization = {{ETSI ISG ISAC}},
  title        = {{GR ISC} 003: Integrated Sensing and Communications ({ISAC}); Use Cases and Deployment Scenarios},
  year         = {2025},
  url = {https://docbox.etsi.org/ISG/ISC/Open}
}

@INPROCEEDINGS{jadoon_icc26,
  author={Jadoon, Muhammad Awais and Robitzsch, Sebastian and Conceicao, Filipe},
  booktitle={2025 IEEE International Conference on Communications Workshops (ICC Workshops)}, 
  title={Dynamic and Resource-Efficient {ISAC} Operations in Sensing-Enabled 6{G} Systems}, 
  year={2025},
  volume={},
  number={},
  pages={2075-2080},
  doi={10.1109/ICCWorkshops67674.2025.11162254}}

@misc{gersing2024architectureproposal6gsystems,
      title={Architecture Proposal for {6G} Systems Integrating Sensing and Communication}, 
      author={Peter Gersing and Mark Doll and Joerg Huschke and Oliver Holschke},
      year={2024},
      eprint={2411.10138},
      archivePrefix={arXiv},
      primaryClass={cs.NI},
      url={https://arxiv.org/abs/2411.10138}, 
}

@misc{hexa-x-ii-d2.3,
    organization = {Hexa-X II},
    title = {Deliverable {D}2.3: Interim overall {6G} system design},
    year = 2024, 
    url = {https://hexa-x-ii.eu/wp-content/uploads/2024/07/Hexa-X-II_D2.3-v1.1.pdf}
}

@INPROCEEDINGS{Moro_icc24,
  author={Moro, Stefano and Linsalata, Francesco and Manzoni, Marco and Magarini, Maurizio and Tebaldini, Stefano},
  booktitle={ICC 2024 - IEEE International Conference on Communications}, 
  title={Exploring {ISAC} Technology for {UAV} {SAR} Imaging}, 
  year={2024},
  volume={},
  number={},
  pages={1582-1587},
  doi={10.1109/ICC51166.2024.10622910}}

@ARTICLE{Liu23_arch,
  author={Liu, Bo and Zhang, Qixun and Jiang, Zheng and Xue, Dongsheng and Xu, Chenlong and Wang, Bowen and She, Xiaoming and Peng, Jinlin},
  journal={China Communications}, 
  title={Architecture for cellular enabled integrated communication and sensing services}, 
  year={2023},
  volume={20},
  number={9},
  pages={59-77},
  doi={10.23919/JCC.fa.2023-0155.202309}}

@INPROCEEDINGS{Lyazidi_24,
  author={Lyazidi, Yazid and Equi, Julia and Shreevastav, Ritesh and Siomina, Iana and Fodor, Gábor},
  booktitle={2024 IEEE Conference on Standards for Communications and Networking (CSCN)}, 
  title={ISAC Architecture and Sensing Topology Switching in 6{G} Cellular Networks}, 
  year={2024},
  volume={},
  number={},
  pages={26-31},
  doi={10.1109/CSCN63874.2024.10849754}}

@INPROCEEDINGS{jadoon_vtc26,
  author={Jadoon, Muhammad Awais and Robitzsch, Sebastian and Babu, Abinaya and Bhatia, Laksh and Conceicao, Filipe},
  booktitle={2025 IEEE 101st Vehicular Technology Conference (VTC2025-Spring)}, 
  title={Enabling {ISAC} in Industry 4.0: Challenges and Opportunities}, 
  year={2025},
  volume={},
  number={},
  pages={1-6},
  doi={10.1109/VTC2025-Spring65109.2025.11174702}}

@INPROCEEDINGS{11230762,
  author={Robitzsch, Sebastian and Bhatia, Laksh and Filis, Konstantinos G. and Petreska, Neda and Bahr, Michael and Martinez, Pablo Picazo and Li, Xi},
  booktitle={2025 IEEE Conference on Standards for Communications and Networking (CSCN)}, 
  title={Architecture Considerations for {ISAC} in 6{G}}, 
  year={2025},
  volume={},
  number={},
  pages={1-5},
  doi={10.1109/CSCN67557.2025.11230762}}
\end{document}